\documentclass{aa}
\usepackage{graphicx}
\usepackage{amssymb}
\usepackage{natbib}

\usepackage{supertabular}
\usepackage{lscape}
\usepackage{longtable}
\usepackage{rotating}%

\def\ssim{\setbox0=\hbox{$\propto$}%
\setbox1=\hbox{$<$}\dimen0=\ht1%
\advance\dimen0by-1.2pt\,\lower.6\dimen0%
\copy0\kern-\wd0\raise.4\dimen0\copy1 \,}

\def\gsim{\setbox0=\hbox{$\propto$}%
\setbox1=\hbox{$>$}\dimen0=\ht1%
\advance\dimen0by-1.2pt\,\lower.6\dimen0%
\copy0\kern-\wd0\raise.4\dimen0\copy1\,}

\def\lambdab{\lambda\mkern-9mu\lower1.2pt\hbox{$\mathchar'26$}}%

\begin{document}
   \title{Yields of rotating stars at solar metallicity}
   \titlerunning{Yields at $Z_{\sun}$ with rotation}


 \author{R. Hirschi \and G. Meynet \and A. Maeder}
 \authorrunning{R. Hirschi et al}
 \institute{Geneva Observatory CH--1290 Sauverny, Switzerland}
\offprints{R. Hirschi \email{Raphael.Hirschi@obs.unige.ch}}

   \date{Received  / Accepted }

\abstract{
We present a new set of stellar yields obtained from 
rotating stellar models at solar metallicity covering the 
massive star range (12--60 $M_{\sun}$). The stellar models were
calculated with
 the latest version of the Geneva stellar evolution code
described in \citet{psn04a}. 
Evolution and nucleosynthesis are in general followed up to silicon
burning.
The yields of our non--rotating models are consistent
with other calculations and differences can be understood in the light
of the treatment of convection and the rate used for
$^{12}$C$(\alpha,\gamma)^{16}$O. This verifies the accuracy of our
calculations and gives a safe basis for studying the effects of rotation
on the yields.
The contributions from stellar winds and
supernova explosions to the stellar yields are presented separately. 
We then add the two contributions to compute the total stellar yields.
Below $\sim 30\,M_{\sun}$, rotation increases the  
total metal yields, $Z$, and in particular the yields of carbon and oxygen 
by a factor of 1.5--2.5.
As a rule of thumb, the yields of a rotating 20 $M_\odot$ star are similar
to the yields of a non--rotating 30 $M_\odot$ star, at least for the light
elements considered in this work.
For very massive stars ($\sim 60\,M_{\sun}$), 
rotation increases the yield of helium but does not significantly 
affect the yields of heavy elements.

 }

\maketitle
%

\section{Introduction}
Stellar yields are a crucial input for galactic chemical evolution. It
is therefore important to update them whenever significant changes
appear in stellar evolution models.
Recent yield calculations at solar metallicity have been conducted by a 
few groups \citep{RHHW02,LC03,TNH96}. 
Over the last ten years, the development of the Geneva evolution 
code has allowed
the study of the evolution of rotating stars until carbon burning. The
models reproduce very well many observational features at various 
metallicities, like
surface enrichments \citep{MM02n}, ratios between red and blue 
supergiants \citep{ROTVII}
and the  population of Wolf--Rayet (WR hereinafter) stars 
\citep{ROTX}. In \citet{psn04a}, we described 
the recent modifications made to the Geneva code
and 
the evolution of our rotating models until silicon burning.
 In this paper, the goal is to calculate
stellar yields for a large initial mass range (12--60 $M_{\sun}$) 
for rotating
stars. In Sect. 2, we briefly present the model
physical ingredients and the calculations. In Sect. 3, we describe the method and
the formulae used
to derive the yields. In Sect. 4, we discuss the wind
contribution to the yields. Then, in Sect. 5,  
we present our supernova (SN) yields of light elements calculated at the 
pre--supernova stage. In Sect. 6, we describe and
analyse the total stellar yields (wind + SN) and 
compare our results with those found in
the literature.
\section{Description of the stellar models}\label{phys}
The computer model used to calculate the stellar models
 is described in detail in \citet{psn04a}.
Convective stability is determined by the 
Schwarzschild criterion. 
Convection is treated as a diffusive process from oxygen burning
onwards.
The overshooting parameter is 0.1 H$_{\rm{P}}$ 
for H and He--burning cores 
and 0 otherwise. On top of the meridional circulation 
and secular shear,
an additional instability induced by rotation, dynamical shear, 
is introduced in the model. 
The reaction rates are taken from the
NACRE \citep{NACRE} compilation for the experimental rates
and from the NACRE website (http://pntpm.ulb.ac.be/nacre.htm) for the
theoretical ones. 

Since mass loss rates are a key ingredient for the yields of massive 
stars, 
we recall here the prescriptions used.
The changes of the mass loss rates, $\dot{M}$, with  
rotation are taken into account as explained in \citet{ROTVI}.
As reference mass loss rates, 
we adopt the mass loss rates of \citet{Vink00,Vink01}
who account for the occurrence of bi--stability
limits which change the wind properties and mass loss rates.
For the domain not covered by these authors
we use the empirical law devised by \citet{Ja88}.
Note that this empirical law, which presents
a discontinuity in the mass flux near the Humphreys--Davidson limit,
implicitly accounts for the mass loss rates of LBV stars. 
For the non--rotating
models, since the empirical values
for the mass loss rates are based on 
stars covering the whole range of rotational velocities, 
we must apply a reduction factor to the empirical rates to make
them correspond to the non--rotating case. The same reduction factor 
was used as in \citet{ROTVII}.
During the Wolf--Rayet phase we use
the mass loss rates by \citet{NuLa00}. 
These mass loss rates,
which account for the clumping effects in the winds,  
are smaller by a factor of 2--3 than the mass loss rates used in our 
previous non--rotating ``enhanced mass loss rate'' stellar grids
\citep{MM94}. 
Wind anisotropy \citep[described in][]{ROTVI} was only taken into account for 
$M \geqslant 40\,M_{\sun}$ since its
effects are only important for very massive stars.

The initial composition of our models is given in Table \ref{inic}.
For a given metallicity $Z$ (in mass fraction), 
the initial helium mass fraction
$Y$ is given by the relation $Y= Y_p + \Delta Y/\Delta Z \cdot Z$, 
where $Y_p$ is the primordial
helium abundance and $\Delta Y/\Delta Z$ the slope of 
the helium--to--metal enrichment law. 
We used the same values as in \citet{ROTVII}
{\it i.e.} $Y_p$ = 0.23 and $\Delta Y/\Delta Z$ = 2.25.
For the solar metallicity, $Z$ = 0.02, we thus have
$X$ = 0.705 and $Y$ = 0.275.
For the mixture of the heavy elements, 
we adopted the same mixture as the one
used to compute the opacity tables for solar composition. For elements 
heavier than Mg, we used the values from \citet{AG89}.

\begin{table}
\caption{Initial abundance (in mass fraction) of the chemical elements.}
\begin{tabular}{l r l r}
\hline
\hline
Element & Mass fraction & Element & Mass fraction \\
\hline \\
 $^{1}$H  & 0.705     & $^{24}$Mg& 5.861D-4 \\
 $^{3}$He & 2.915D-5  & $^{25}$Mg& 7.70D-5  \\
 $^{4}$He & 0.275     & $^{26}$Mg& 8.84D-5  \\
 $^{12}$C & 3.4245D-3 & $^{28}$Si& 6.5301D-4\\
 $^{13}$C & 4.12D-5   & $^{32}$S & 3.9581D-4\\ 
 $^{14}$N & 1.0589D-3 & $^{36}$Ar& 7.7402D-5\\ 
 $^{15}$N & 4.1D-6    & $^{40}$Ca& 5.9898D-5\\ 
 $^{16}$O & 9.6195D-3 & $^{44}$Ti& 0        \\ 
 $^{17}$O & 3.9D-6    & $^{48}$Cr& 0        \\ 
 $^{18}$O & 2.21D-5   & $^{52}$Fe& 0        \\ 
 $^{20}$Ne& 1.8222D-3 & $^{56}$Ni& 0        \\ 
 $^{22}$Ne& 1.466D-4  &          &          \\
\hline
\end{tabular}
\label{inic}
\end{table}

We calculated stellar models with initial masses of 12, 15, 20, 25, 40 and 60 $M_{\sun}$
 at solar metallicity, with initial rotation velocities 
of 0 and 300 km\,s$^{-1}$. The value of the initial velocity
corresponds to an average velocity of about 220\,km\,s$^{-1}$ on the Main
Sequence (MS) which is
very close to the average observed value \citep[see for instance][]{FU82}. 
The calculations start at the ZAMS.
Except for the 12 $M_{\sun}$ models, the rotating models were computed until
the end of core silicon (Si) burning and their non--rotating counterparts
until the end of shell Si--burning. 
For the non--rotating 12 $M_{\sun}$ star, neon (Ne) burning starts at
a fraction of a solar mass away from the centre but does not reach the centre and the 
calculations stop there. For the rotating 12 $M_{\sun}$ star, the model 
ends after oxygen (O) burning. The evolution of the models is described in \citet{psn04a}.

\section{Yield calculations}\label{def}
In this paper, we calculated separately the yield contributions 
from stellar winds and the SN explosion.
The wind contribution from a star of initial mass, $m$, 
to the stellar yield of an element $i$ is:
\begin{equation}\label{wdef}  
mp^{\rm{wind}}_{im}= \int_{0}^{\tau(m)} \dot{M}(m,t)[X_{i}^S(m,t)-X_{i}^{0}]\,dt
\end{equation}where 
$\tau(m)$ is the final age of the star, 
 $\dot{M}(m,t)$ the mass loss rate when the age of the star is equal to
 $t$,
 $X_{i}^S(m,t)$ the surface abundance in mass fraction of element $i$ 
and $X_{i}^{0}$ its initial mass fraction 
(see Table \ref{inic}). 
Mass loss occurs mainly during hydrogen (H) and helium (He) burnings. 
Indeed, the advanced
stages of the hydrostatic evolution are so short in time that
only a negligible amount of mass is lost during these phases. 

In order to calculate the SN explosion contribution to stellar yields 
of all the chemical elements, 
one needs to model the complete evolution of the star from the ZAMS 
up to and including
the SN explosion. However, elements
lighter than neon are marginally modified by explosive
nucleosynthesis \citep{CL03,TNH96} and are mainly determined by the hydrostatic
evolution while elements between neon and silicon are produced both 
hydrostatically and explosively. In this work, we calculate SN yields
at the end of core Si--burning. We therefore 
present these yields as pre--SN yields. 
The pre--SN contribution from a star of initial mass, $m$, 
to the stellar yield of an element $i$ is:
\begin{equation}\label{sndef}
mp^{\rm{pre-SN}}_{im}= \int_{m(\rm{rem})}^{m(\tau)} [X_{i}(m_r)-X_{i}^{0}]\,dm_r  
\end{equation}where 
$m(\rm{rem})$ is the remnant mass,
$m(\tau)$ the final stellar mass,
$X_{i}^{0}$ the initial abundance in mass fraction of element $i$
and $X_{i}(m_r)$ the final abundance in mass fraction at the lagrangian
mass coordinate, $m_r$. 

The remnant mass in Eq. \ref{sndef} corresponds to
the final baryonic remnant mass that includes
fallback that may occur after the SN explosion.
The exact determination of the remnant mass would again require
the simulation of the core collapse and 
SN explosion, which is not within the scope of this paper. 
Even if we had done
the simulation, the remnant mass would still be a free parameter 
because most explosion models
still struggle to reproduce explosions 
\citep[ and references therein]{J03}. 
Nevertheless, the latest multi--D simulations \citep{J04} show that
low modes of the convective instability may help produce an explosion.
When comparing models to observations, the
 remnant mass is usually chosen so that the amount of radioactive
$^{56}$Ni ejected by the star corresponds to the value determined from 
the observed light curve. 
So far, mostly 1D models are used for explosive nucleosynthesis but
a few groups have developed multi--D models 
\citep[see][]{TKM04,MN03}.  
Multi--D effects like mixing and asymmetry might play a role in
determining the mass
cut if some iron group elements are mixed with oxygen-- or silicon--rich
layers. 
\begin{table}
\caption{Initial mass (column 1) and initial rotation 
velocity [km\,s$^{-1}$] (2), final mass (3), 
masses of the helium (4), carbon--oxygen (5) cores, 
the remnant mass (6) and lifetimes [Myr] (7) for solar metallicity 
models. All masses are in
solar mass units. An "A" in the second column
means that wind anisotropy was taken into account.}
\begin{tabular}{r r r r r r r r}
\hline \hline 
$M_{\rm{ini}}$ & $\upsilon_{\rm{ini}}$ & $M_{\rm{final}}$ & $M_{\alpha}$ 
& $M_{\rm{CO}}$ & $M_{\rm{rem}}$ & $\tau_{\rm{life}}$ \\ 
\hline
 12 & 0    & 11.524 &  3.141 &  1.803 & 1.342 &  18.01 \\    
 12 & 300  & 10.199 &  3.877 &  2.258 & 1.462 &  21.89 \\   
 15 & 0    & 13.232 &  4.211 &  2.441 & 1.510 &  12.84 \\   
 15 & 300  & 10.316 &  5.677 &  3.756 & 1.849 &  15.64 \\   
 20 & 0    & 15.694 &  6.265 &  4.134 & 1.945 &   8.93 \\   
 20 & 300  &  8.763 &  8.654 &  6.590 & 2.566 &  10.96 \\  
 25 & 0    & 16.002 &  8.498 &  6.272 & 2.486 &   7.32 \\   
 25 & 300  & 10.042 & 10.042 &  8.630 & 3.058 &   8.67 \\   
 40 & 0    & 13.967 & 13.967 & 12.699 & 4.021 &   5.05 \\   
 40 & 300A & 12.646 & 12.646 & 11.989 & 3.853 &   5.97 \\   
 60 & 0    & 14.524 & 14.524 & 13.891 & 4.303 &   4.02 \\  
 60 & 300A & 14.574 & 14.574 & 13.955 & 4.323 &   4.69 \\
\hline
\end{tabular}
\label{table1}
\end{table}

In this work, we used the relation between
$M_{\rm{CO}}$ and the remnant mass described in \citet{AM92}. 
The resulting remnant mass as well as the major characteristics of the
models are given in Table \ref{table1}. The determination of 
$M_{\alpha}$ and $M_{\rm{CO}}$ is described in \citet{psn04a}. 

We do not follow $^{22}$Ne after He--burning 
and have to apply a
special criterion to calculate its pre--SN yield. 
During He--burning,
$^{22}$Ne is
produced by $^{18}\rm{O}(\alpha,\gamma)$ and 
destroyed by an
$\alpha$--captures which create $^{25}$Mg or $^{26}$Mg. 
$^{22}$Ne is totally destroyed by
C--burning. We therefore consider $^{22}$Ne abundance to be null inside 
of the C--burning shell. Numerically, this is done when the mass fraction of
$^{4}$He is less than $10^{-4}$ and that of $^{12}$C is less than 0.1. 

Once both the wind and pre--SN contributions are calculated, 
the total stellar yield of an element $i$ from a star of initial 
mass, $m$, is:
\begin{equation}\label{ydef}
mp^{\rm{tot}}_{im} = mp^{\rm{pre-SN}}_{im} + mp^{\rm{wind}}_{im}
\end{equation}$mp^{\rm{tot}}_{im}$ corresponds to the amount of element 
$i$ newly synthesised and ejected by a star
of initial mass $m$ \citep[see][]{AM92}. 

Other authors give their results in ejected masses, $EM$:
\begin{equation}\label{emdef}
EM_{im}= \int_{0}^{\tau(m)} \dot{M}\,X_{i}^S(m,t)\,dt 
+ \int_{m(\rm{rem})}^{m(\tau)}\,X_{i}(m_r)\,dm_r
\end{equation} and production factors (PF) \citep[see][]{WLW95}:
\begin{equation}\label{fdef}
f_{im}= EM_{im}/[X_{i}^{0}(m-m(\rm{rem}))] 
\end{equation}We also give our final results as ejected masses in order 
to compare our results with
the recent literature.

\section{Stellar wind contribution}\label{WY}
\begin{table*}
\caption{
Initial mass and velocity and 
{\bf stellar wind contribution to the yields} ($mp^{\rm{wind}}_{im}$). 
All masses and yields are in
solar mass units and velocities are in km\,s$^{-1}$.
"A" in column 1 means wind anisotropy has been included in the model. Z is the total
metal content and is defined by: Z$=1-X_{\rm{^1 H}}-X_{\rm{^3 He}}- 
X_{\rm{^4 He}}$.}
\begin{tabular}{l r r r r r r r r r r r}
\hline
\hline \\
$M_{\rm{ini}}$, $\upsilon_{\rm{ini}}$ 
            & $^{3}$He & $^4$He & $^{12}$C & $^{13}$C & $^{14}$N & $^{16}$O & $^{17}$O & $^{18}$O & $^{20}$Ne & $^{22}$Ne & Z \\
\hline
 12,    0   & -2.49E-6 &  1.55E-2 & -4.80E-4 &  2.53E-5 &  9.39E-4 & -4.62E-4 &  1.43E-6 & -2.57E-6 & -9.51E-8 &  0        &  0        \\
 12,  300   & -1.59E-5 &  1.20E-1 & -2.54E-3 &  2.09E-4 &  5.18E-3 & -2.78E-3 &  7.90E-6 & -1.36E-5 & -3.60E-7 &  0        &  0        \\
 15,    0   & -4.15E-6 &  9.49E-3 & -8.01E-4 &  1.54E-4 &  1.02E-3 & -2.87E-4 &  9.20E-7 & -2.92E-6 & -3.54E-7 &  0        &  0        \\
 15,  300   & -5.23E-5 &  3.25E-1 & -6.73E-3 &  5.78E-4 &  1.39E-2 & -7.63E-3 &  1.63E-5 & -3.64E-5 & -9.37E-7 &  0        &  0        \\
 20,    0   & -1.06E-5 &  5.21E-2 & -1.21E-3 &  1.93E-4 &  2.46E-3 & -1.43E-3 &  1.18E-6 & -6.28E-6 & -8.61E-7 &  0        &  0        \\
 20,  300   & -1.56E-4 &  1.27E+0 & -1.73E-2 &  1.22E-3 &  4.30E-2 & -2.75E-2 &  2.03E-5 & -1.01E-4 & -2.25E-6 &  0        &  0        \\
 25,    0   & -6.02E-5 &  3.95E-1 & -6.39E-3 &  4.38E-4 &  1.64E-2 & -1.07E-2 &  3.39E-6 & -3.88E-5 & -1.80E-6 &  0        &  0        \\
 25,  300   & -2.47E-4 &  2.97E+0 & -2.52E-2 &  1.22E-3 &  7.94E-2 & -5.48E-2 &  1.03E-5 & -1.68E-4 & -2.99E-6 &  2.72E-4  &  0        \\
 40,    0   & -4.16E-4 &  4.65E+0 & -4.48E-2 &  7.11E-4 &  1.45E-1 & -1.07E-1 & -1.10E-5 & -3.04E-4 & -5.18E-6 &  0        &  0        \\
 40,  300A  & -5.83E-4 &  7.97E+0 &  1.60E+0 &  5.42E-4 &  1.73E-1 &  3.34E-1 & -3.93E-5 & -4.11E-4 &  1.76E-6 &  7.35E-2  &  2.18     \\
 60,    0   & -8.33E-4 &  9.41E+0 &  2.52E+0 &  4.15E-4 &  2.37E-1 &  4.45E-1 & -6.24E-5 & -6.32E-4 &  2.46E-6 &  1.28E-1  &  3.33     \\
 60,  300A  & -1.07E-3 &  1.52E+1 &  3.01E+0 &  3.12E-4 &  3.09E-1 &  3.99E-1 & -9.85E-5 & -8.10E-4 &  3.39E-6 &  1.67E-1  &  3.89     \\
\hline
\end{tabular}
\label{tw}
\end{table*}
Before we discuss the wind contribution to the stellar yields, it is
instructive to look at the final masses given in Table \ref{table1} 
\citep[see also Fig. 16 in][]{psn04a}.
We see that, below 25 $M_{\sun}$, the rotating models lose significantly
more mass. 
This is due to the fact that rotation enhances mass loss
 and favours the evolution into the red supergiant phase at an early
 stage during the core He--burning phase \citep[see for example][]{MM00}.
For WR stars ($M \gtrsim 30 \,M_{\sun}$), the new prescription by 
\citet{NuLa00}, including the
effects of clumping in the winds, 
results in mass loss rates that are a factor of two to
three smaller than the rates from \citet{La89}. 
The final masses of very massive stars ($\gtrsim 60\,M_{\sun}$) are therefore
never small enough to produce neutron stars. 
We therefore expect the
same outcome (BH formation) for the very massive stars
as for the stars with masses around 40 $M_{\sun}$ at solar metallicity.

The wind contribution to the stellar yields is presented in Table 
\ref{tw}. 
The H--burning products (main elements are $^4$He and $^{14}$N) are ejected by
stellar winds in the entire massive star range. Nevertheless, in absolute value,
the quantities ejected by very massive stars ($M \gtrsim 40 \,M_{\sun}$) are much
larger. 
These yields are larger in rotating models. This is due to both the increase of
mixing and mass loss by rotation. For $M \lesssim 40 \,M_{\sun}$,
the dominant effect is the diffusion of H--burning products in the envelope of the star due to
rotational mixing. For more massive stars ($M \gtrsim 40 \,M_{\sun}$), 
the mass loss effect is dominant.

The He--burning products are produced deeper in the star. They are 
therefore ejected only by WR star winds. Since
the new mass loss rates are reduced by a factor of two to three (see Sect.
\ref{phys}), the yields from the winds in $^{12}$C are much smaller 
for the present
WR stellar models compared to the results obtained in \citet{AM92}. 
As is shown below, the
pre--SN contribution to the yields of $^{12}$C are larger in the present
calculation and, as a matter of fact, 
the new $^{12}$C total yields are larger than in \citet{AM92}. 
In general, the yields for rotating stars are larger than for
non--rotating ones due to the extra
mass loss and mixing. For very massive stars 
($M \gtrsim 60 \,M_{\sun}$), 
the situation is reversed for He--burning products because of 
the different mass loss history.
Indeed, rotating stars enter into the WR regime in the course of the
main sequence (MS). In particular, the long time spent in the WNL phase
\citep[WN star showing hydrogen at its surface,][]{ROTX} 
results in the ejection of large amounts of
H--burning products. Therefore, the star enters the WC phase with a
smaller total mass and fewer He--burning products are ejected by winds
(the mass loss being proportional to the actual mass of the star).

Since $^{16}$O is produced even deeper in the star, the present
contribution by winds to this yield are even smaller. 
$^{12}$C constituting the largest fraction of ejected metals, 
the conclusion for the wind contribution to the total metallic yield, Z,
is the same as for $^{12}$C.

\begin{figure*}[!tbp]
\centering
\includegraphics[width=7.5cm]{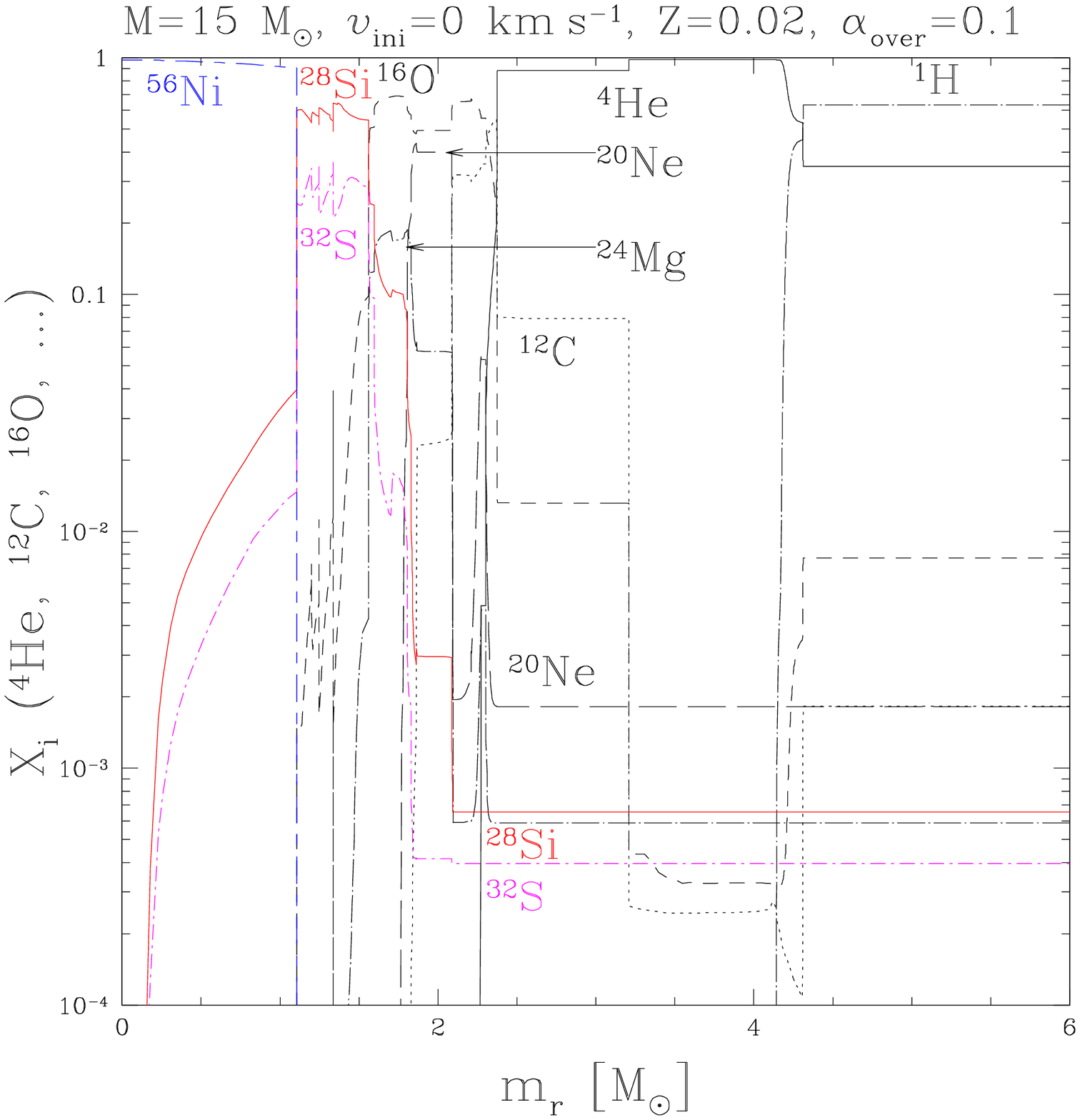}\includegraphics[width=7.5cm]{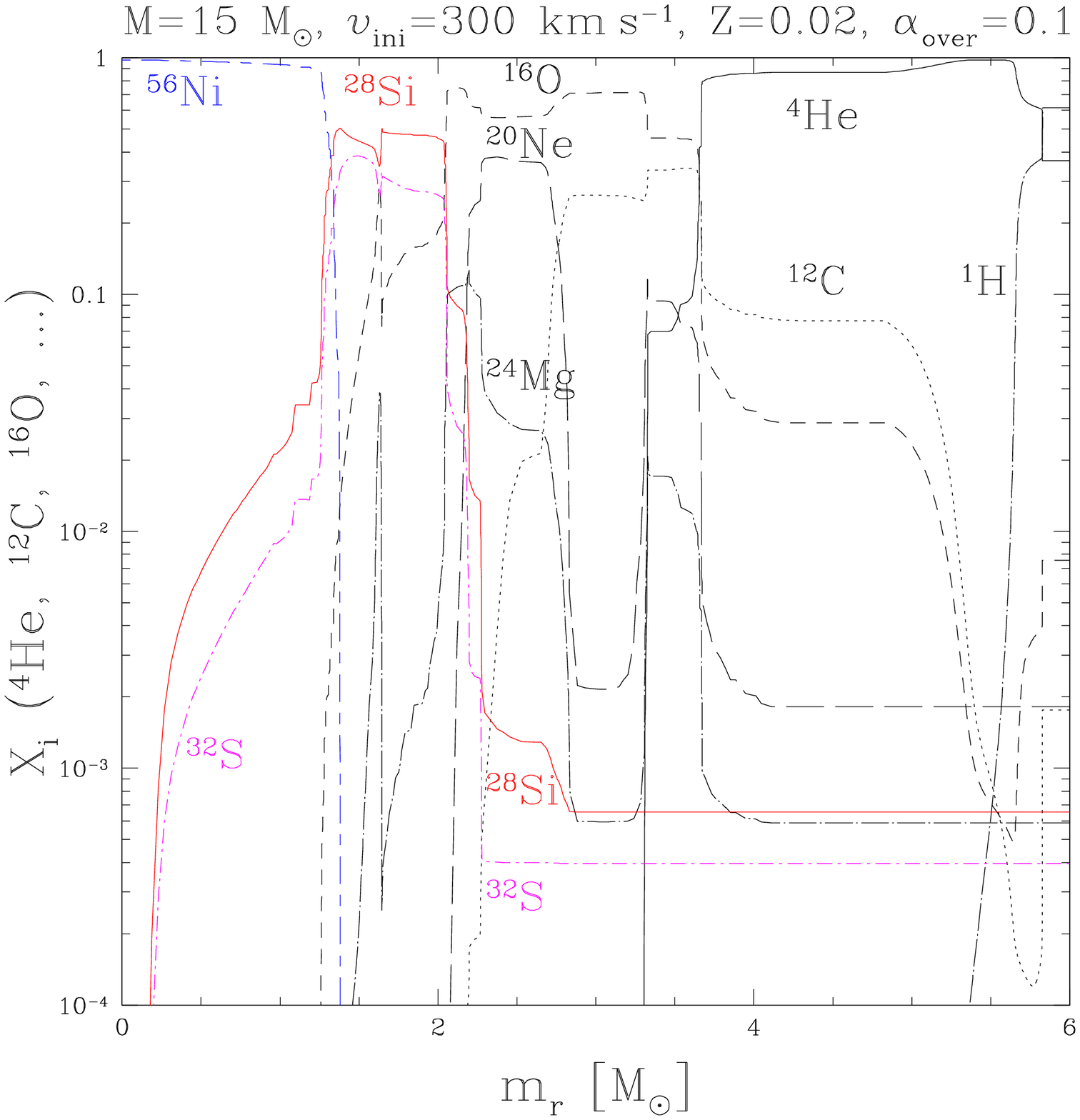}
\includegraphics[width=7.5cm]{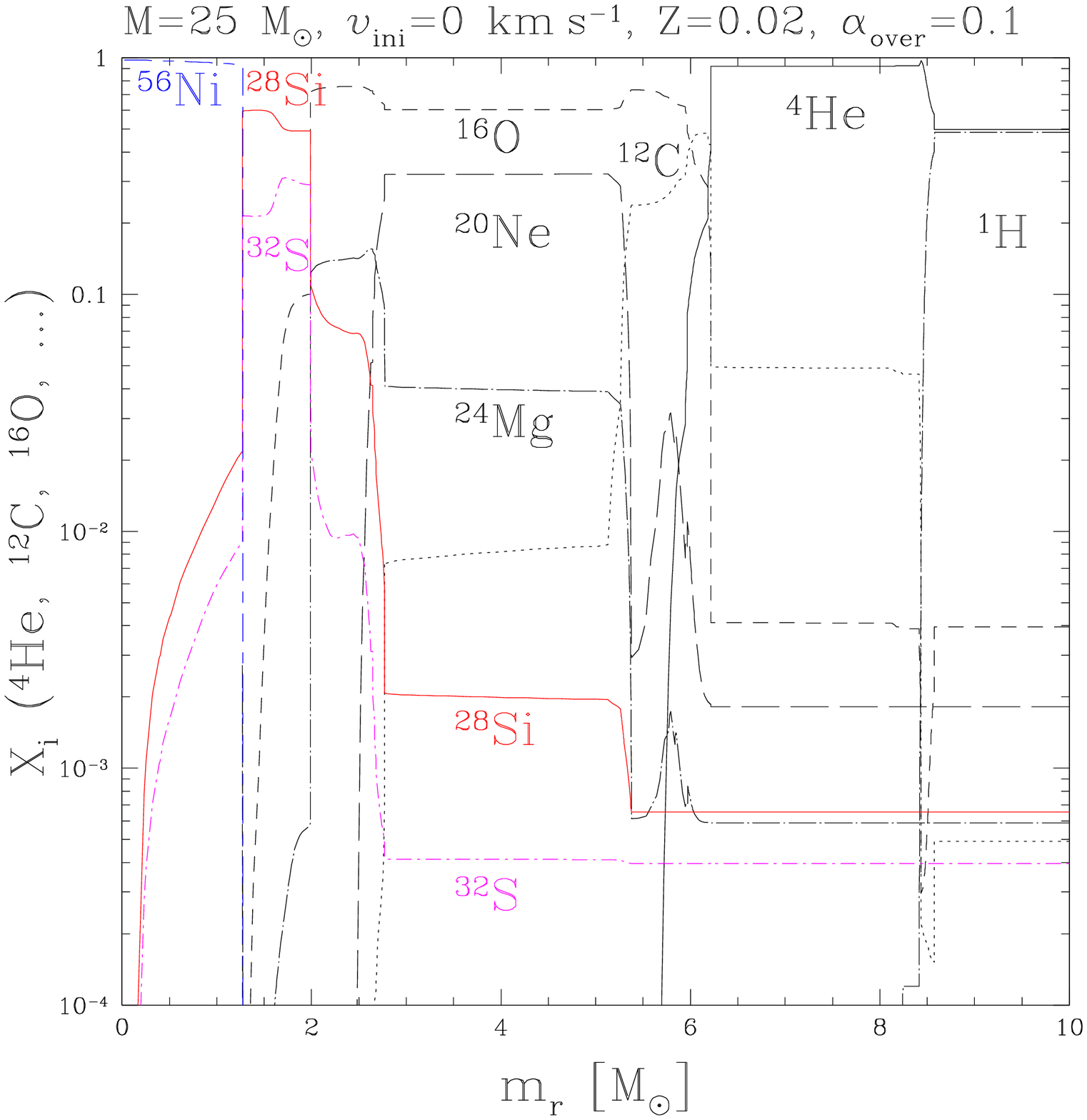}\includegraphics[width=7.5cm]{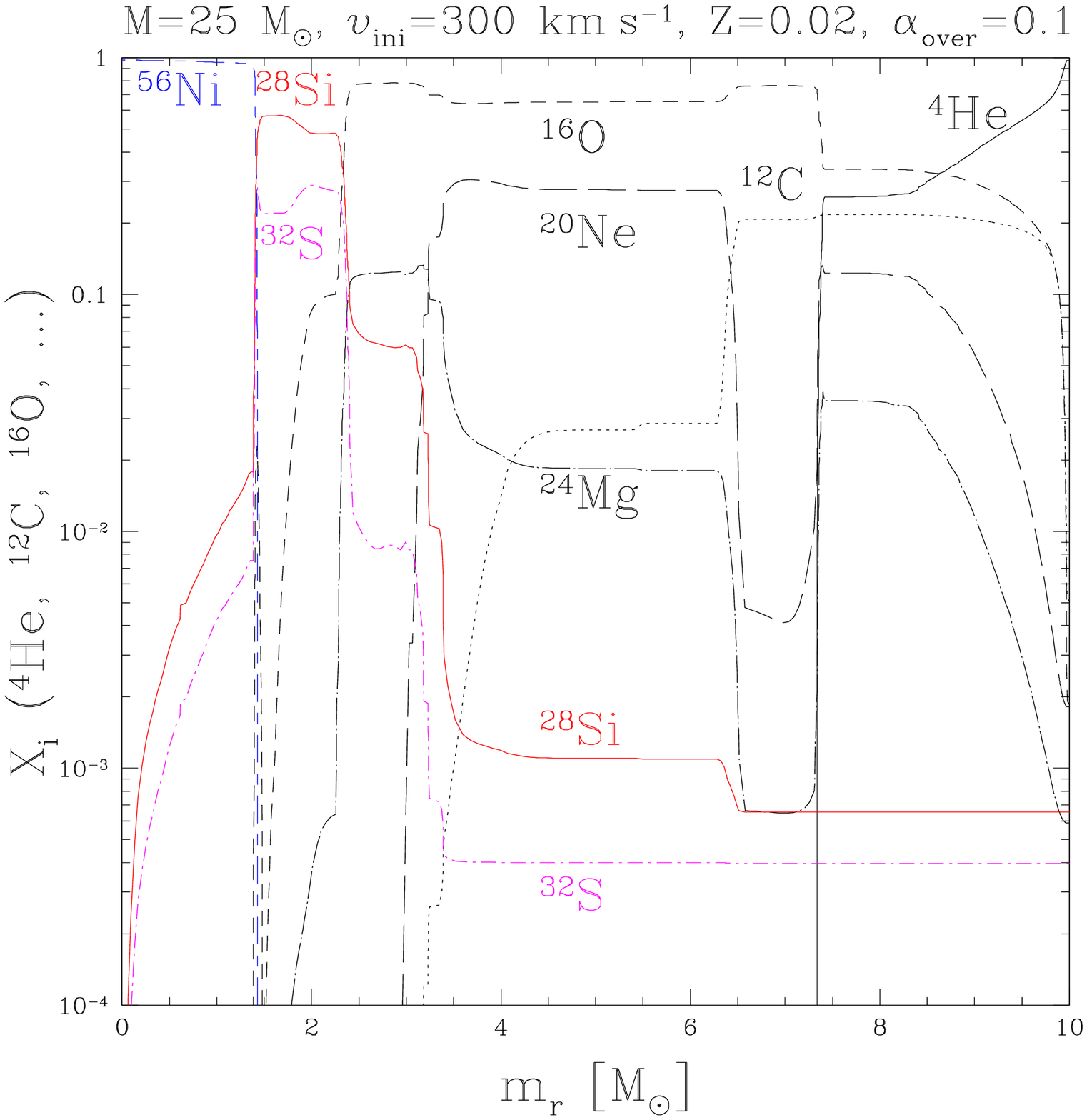}
\includegraphics[width=7.5cm]{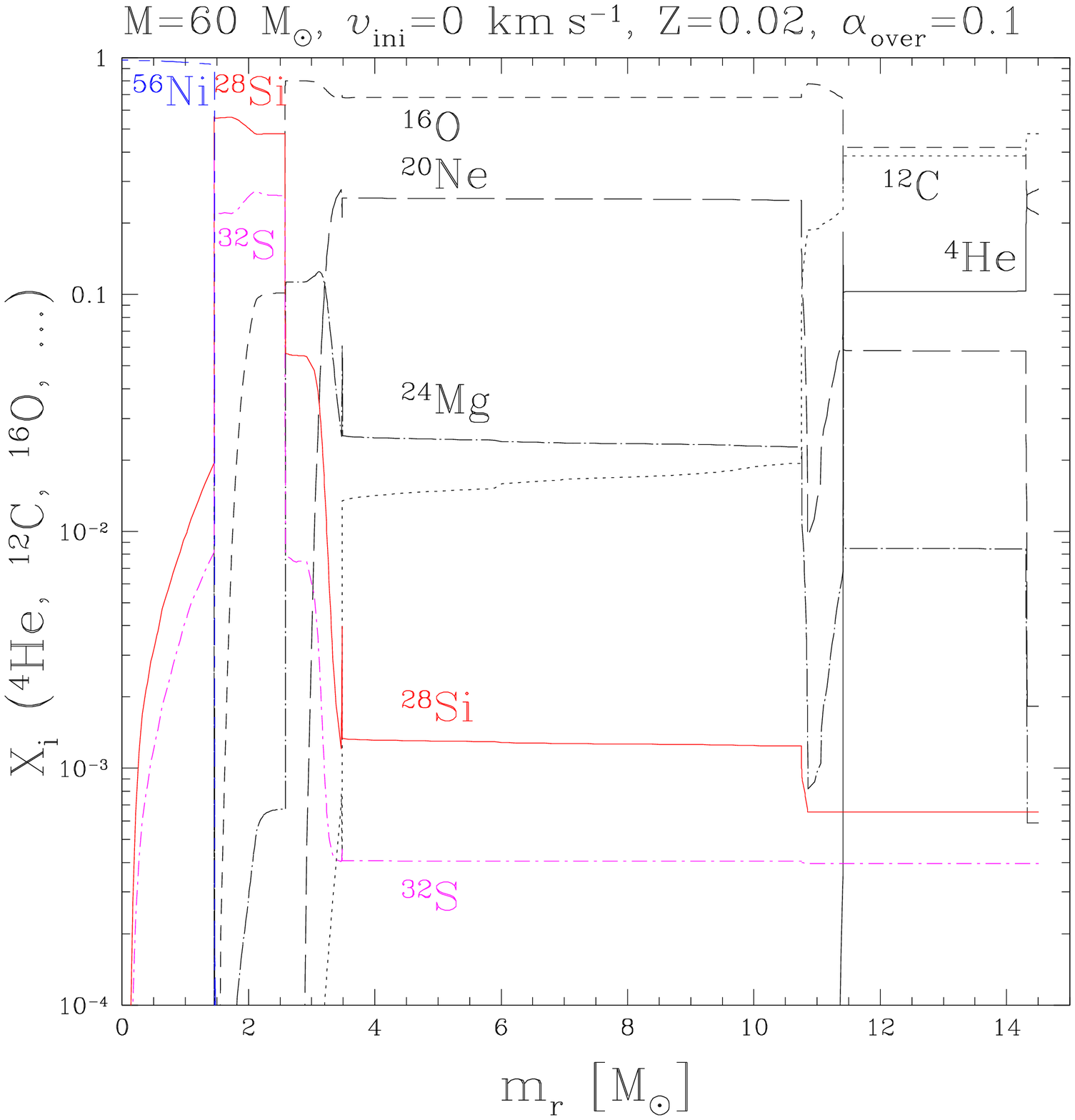}\includegraphics[width=7.5cm]{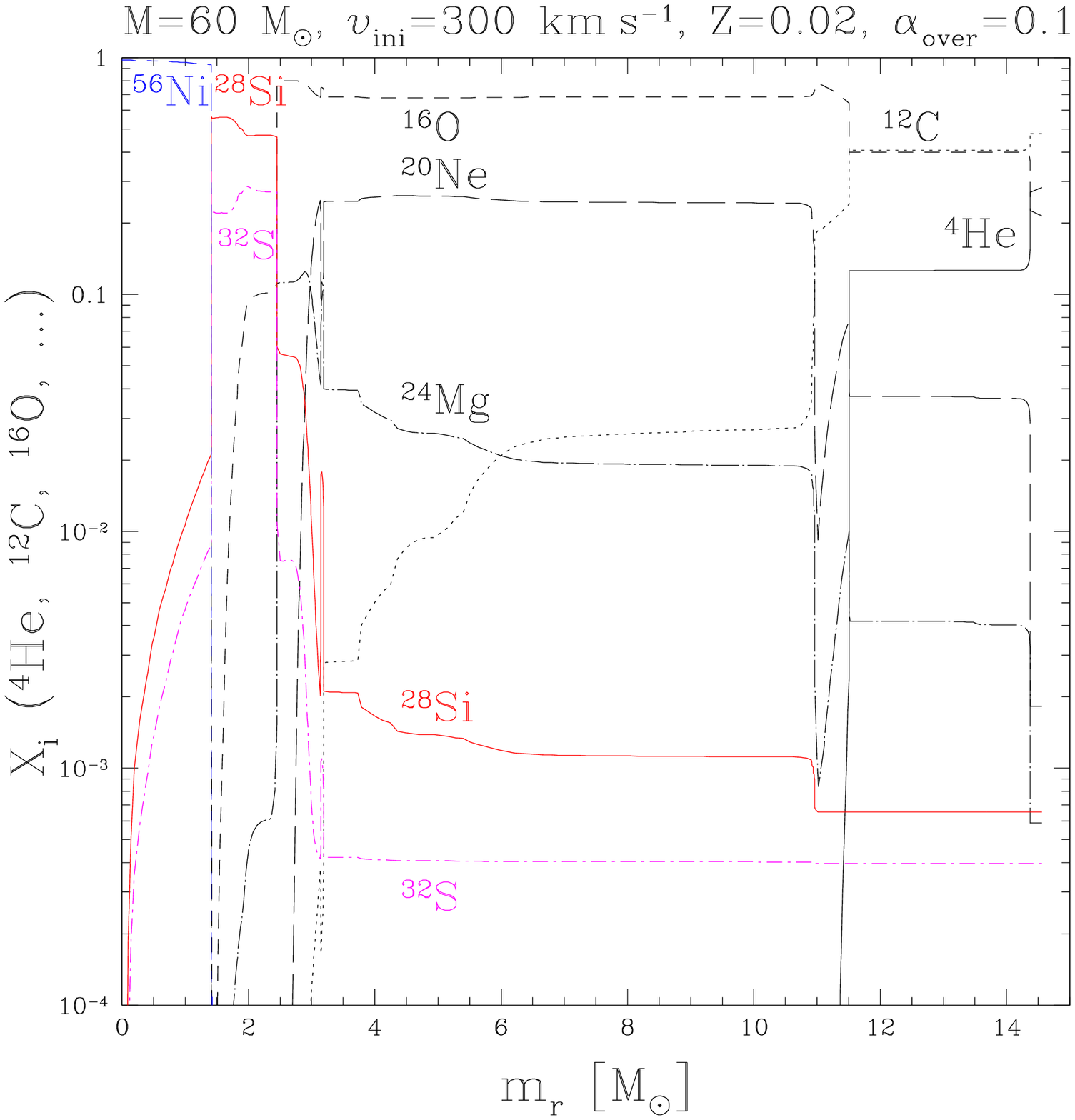}
\caption{Abundance profile at the end of core silicon
burning for the non--rotating (left) and rotating
 (right) 
 15 (top), 25 (middle) and 60 (bottom) $M_{\sun}$ models.}
\label{abm152560}
\end{figure*}
\section{Pre--SN contribution}
\begin{table*}
\caption{
Initial mass and velocity and 
{\bf pre--SN contribution to the yields} ($mp^{\rm{pre-SN}}_{im}$) 
of solar metallicity models. All masses and yields are in
solar mass units and velocities are in km\,s$^{-1}$.
"A" in column 2 means wind anisotropy has been included in the model.
Z is the total
metal content and is defined by: Z$=1-X_{\rm{^1 H}}-X_{\rm{^3 He}}- 
X_{\rm{^4 He}}$.}
\begin{tabular}{l l r r r r r r r r r}
\hline
\hline \\
$M_{\rm{ini}}$ & $\upsilon_{\rm{ini}}$ & $^{3}$He & $^4$He & $^{12}$C & $^{13}$C & $^{14}$N & $^{16}$O & $^{17}$O & $^{18}$O & Z \\
\hline
 12 &    0   &  -1.22E-4 &  1.19E+0 &  8.74E-2 &  5.83E-4 &  3.59E-2 &  2.07E-1 &  3.45E-5 &  1.62E-4 &  4.57E-1     \\
 12 &  300   &  -1.45E-4 &  1.48E+0 &  1.66E-1 &  8.72E-4 &  3.52E-2 &  3.94E-1 &  2.68E-5 &  1.69E-3 &  7.97E-1     \\
 15 &    0   &  -1.87E-4 &  1.67E+0 &  1.54E-1 &  5.51E-4 &  4.23E-2 &  4.34E-1 &  1.51E-5 &  2.76E-3 &  9.19E-1     \\
 15 &  300   &  -1.75E-4 &  1.24E+0 &  3.68E-1 &  4.90E-4 &  2.13E-2 &  1.02E+0 &  3.87E-6 &  3.76E-3 &  1.92E+0     \\
 20 &    0   &  -2.88E-4 &  2.03E+0 &  2.17E-1 &  4.39E-4 &  4.72E-2 &  1.20E+0 &  7.76E-6 &  3.92E-3 &  2.17E+0     \\
 20 &  300   &  -1.81E-4 &  3.47E-1 &  4.50E-1 & -2.09E-4 &  2.81E-4 &  2.60E+0 & -2.30E-5 & -9.54E-5 &  3.98E+0     \\
 25 &    0   &  -3.74E-4 &  2.18E+0 &  3.74E-1 & -2.51E-5 &  5.76E-2 &  2.18E+0 & -6.73E-6 & -1.74E-4 &  3.74E+0     \\
 25 &  300   &  -2.04E-4 & -8.31E-1 &  7.62E-1 & -2.65E-4 & -5.60E-3 &  3.61E+0 & -2.72E-5 &  4.34E-4 &  5.75E+0     \\
 40 &    0   &  -2.90E-4 & -1.64E+0 &  6.94E-1 & -4.10E-4 & -1.05E-2 &  6.23E+0 & -3.88E-5 & -2.20E-4 &  8.66E+0     \\
 40 &  300A  &  -2.56E-4 & -2.08E+0 &  1.51E+0 & -3.62E-4 & -9.31E-3 &  5.03E+0 & -3.43E-5 & -1.94E-4 &  8.28E+0     \\
 60 &    0   &  -2.98E-4 & -2.45E+0 &  1.42E+0 & -4.21E-4 & -1.08E-2 &  6.03E+0 & -3.99E-5 & -2.26E-4 &  9.66E+0     \\
 60 &  300A  &  -2.99E-4 & -2.40E+0 &  1.49E+0 & -4.22E-4 & -1.09E-2 &  6.01E+0 & -4.00E-5 & -2.27E-4 &  9.63E+0     \\
\hline
\end{tabular}
\label{tsn}
\end{table*}
\begin{table}
\caption{{\bf Pre--SN contribution to the yields} ($mp^{\rm{pre-SN}}_{im}$) 
of solar metallicity models. Continuation of Table \ref{tsn}. 
Note that $^{20}$Ne yields are an upper limit 
and may be reduced by Ne--explosive burning and that $^{24}$Mg yields may
also be modified by neon and oxygen explosive burnings. See discussion in Sect. \ref{yao}.}
\begin{tabular}{l l r r r}
\hline
\hline \\
$M_{\rm{ini}}$ & $\upsilon_{\rm{ini}}$ & ($^{20}$Ne) & $^{22}$Ne & ($^{24}$Mg) \\
\hline
 12 &    0   &  1.05E-1 &  6.80E-3 &  6.75E-3      \\
 12 &  300   &  1.58E-1 &  1.55E-2 &  1.36E-2      \\
 15 &    0   &  1.10E-1 &  1.60E-2 &  6.06E-2      \\
 15 &  300   &  2.26E-1 &  3.33E-2 &  4.27E-2      \\
 20 &    0   &  4.83E-1 &  3.64E-2 &  1.28E-1      \\
 20 &  300   &  6.80E-1 &  4.26E-2 &  1.19E-1      \\
 25 &    0   &  8.41E-1 &  5.22E-2 &  1.38E-1      \\
 25 &  300   &  1.08E+0 &  2.24E-2 &  1.48E-1      \\
 40 &    0   &  1.36E+0 &  5.58E-3 &  1.52E-1      \\
 40 &  300A  &  1.42E+0 &  6.93E-3 &  1.20E-1      \\
 60 &    0   &  1.81E+0 &  6.58E-3 &  1.73E-1      \\
 60 &  300A  &  1.75E+0 &  7.60E-3 &  1.44E-1      \\
\hline
\end{tabular}
\label{tsnb}
\end{table}
\begin{table*}
\caption{
Initial mass and velocity and 
{\bf total stellar yields} ($mp^{\rm{pre-SN}}_{im} + mp^{\rm{wind}}_{im}$) 
 of solar metallicity models. All masses and yields are in
solar mass units and velocities are in km\,s$^{-1}$.
"A" means wind anisotropy has been included in the model. 
Z is the total
metal content and is defined by: Z$=1-X_{\rm{^1 H}}-X_{\rm{^3 He}}- 
X_{\rm{^4 He}}$. These are the yields to be used for chemical evolution 
models using Eq. 2 from \citep{AM92}.}
\begin{tabular}{l l r r r r r r r r r}
\hline
\hline \\
$M_{\rm{ini}}$ & $\upsilon_{\rm{ini}}$ & $^{3}$He & $^4$He & $^{12}$C & $^{13}$C & $^{14}$N & $^{16}$O & $^{17}$O & $^{18}$O & Z \\
\hline
 12 &    0   &  -1.24E-4 &  1.20E+0 &  8.69E-2 &  6.08E-4 &  3.68E-2 &  2.07E-1 &  3.59E-5 &  1.60E-4 &  4.57E-1  \\
 12 &  300   &  -1.61E-4 &  1.60E+0 &  1.63E-1 &  1.08E-3 &  4.04E-2 &  3.92E-1 &  3.47E-5 &  1.67E-3 &  7.97E-1  \\
 15 &    0   &  -1.91E-4 &  1.67E+0 &  1.53E-1 &  7.05E-4 &  4.33E-2 &  4.34E-1 &  1.60E-5 &  2.76E-3 &  9.19E-1  \\
 15 &  300   &  -2.27E-4 &  1.57E+0 &  3.61E-1 &  1.07E-3 &  3.52E-2 &  1.01E+0 &  2.02E-5 &  3.72E-3 &  1.92E+0  \\
 20 &    0   &  -2.98E-4 &  2.08E+0 &  2.16E-1 &  6.32E-4 &  4.96E-2 &  1.20E+0 &  8.94E-6 &  3.91E-3 &  2.17E+0  \\
 20 &  300   &  -3.36E-4 &  1.62E+0 &  4.33E-1 &  1.01E-3 &  4.33E-2 &  2.57E+0 & -2.75E-6 & -1.96E-4 &  3.98E+0  \\
 25 &    0   &  -4.34E-4 &  2.57E+0 &  3.68E-1 &  4.13E-4 &  7.40E-2 &  2.17E+0 & -3.34E-6 & -2.12E-4 &  3.74E+0  \\
 25 &  300   &  -4.51E-4 &  2.14E+0 &  7.37E-1 &  9.57E-4 &  7.38E-2 &  3.55E+0 & -1.69E-5 &  2.66E-4 &  5.75E+0  \\
 40 &    0   &  -7.06E-4 &  3.01E+0 &  6.49E-1 &  3.01E-4 &  1.35E-1 &  6.12E+0 & -4.98E-5 & -5.24E-4 &  8.65E+0  \\
 40 &  300A  &  -8.39E-4 &  5.89E+0 &  3.11E+0 &  1.80E-4 &  1.63E-1 &  5.36E+0 & -7.35E-5 & -6.05E-4 &  1.05E+1  \\
 60 &    0   &  -1.13E-3 &  6.96E+0 &  3.94E+0 & -6.23E-6 &  2.26E-1 &  6.47E+0 & -1.02E-4 & -8.58E-4 &  1.30E+1  \\
 60 &  300A  &  -1.37E-3 &  1.28E+1 &  4.51E+0 & -1.11E-4 &  2.98E-1 &  6.41E+0 & -1.38E-4 & -1.04E-3 &  1.35E+1  \\
\hline
\end{tabular}
\label{ytot}
\end{table*}
As said above, our pre--SN yields, $mp^{\rm{pre-SN}}_{im}$, were calculated at
 the end of core Si--burning using the remnant mass, $M_{\rm{rem}}$, 
 given in Table \ref{table1}. 
We therefore concentrate on yields of light
elements which depend mainly on the evolution prior to core Si--burning. 

Before discussing the pre--SN yields, it is interesting to look at the abundance
profiles at the pre--SN stage presented in Fig. \ref{abm152560} and at the
size of helium and carbon-oxygen cores given in Table \ref{table1}.
The core sizes are clearly increased due to rotational mixing. We also see that
as the initial mass of the model increases, the core masses get closer and closer
to the final mass of the star. $M_{\alpha}$ reaches the final mass of the star
when $M_{\rm{ini}} \gtrsim 40\,M_\odot$ for non--rotating models and 
when $M_{\rm{ini}} \gtrsim 25\,M_\odot$ 
for rotating models. $M_{\rm{CO}}$ becomes close to the final mass
for both rotating and non--rotating models for $M_{\rm{ini}} \gtrsim 40\,M_\odot$.

The pre--SN yields are presented in Tables  \ref{tsn} and \ref{tsnb}. 
One surprising result in Table
\ref{tsn} is the negative pre--SN yields of $^4$He (and of $^{14}$N) for WR stars. 
This is simply due to
the definition of stellar yields, in which the initial composition is deducted
from the final one. As said above, $M_{\rm{CO}}$ becomes close to the final mass
for $M_{\rm{ini}} \gtrsim 40\,M_\odot$. Since
the CO core is free of helium, it is then understandable that the pre--SN yields of
$^4$He for WR stars is negative.

\subsection{Carbon, oxygen and metallic yields}\label{yov}
If mixing is
dominant ($M \lesssim 30\,M_{\sun}$), the larger the initial mass, the
larger the metallic yields (because the various cores become larger). 
Rotation increases the core sizes by extra mixing and therefore the 
total metallic
yields are larger for rotating models. 
Overshooting also plays a role in the core sizes. The larger the
overshooting parameter, the larger the cores and the larger the yields. 
If we compare our rotating and non--rotating models, we see that the
pre--SN total metallic yields
and $^{12}$C and $^{16}$O yields in particular are enhanced by rotation 
by a factor 1.5--2.5 below 30\,$M_{\sun}$.

For very massive stars ($M \gtrsim
30\,M_{\sun}$), the higher the mass loss, the smaller the final mass 
and the total metallic yields.
The same explanations work well in general for carbon and oxygen. 
\section{Total stellar yields}
\begin{figure*}[!tbp]
\centering
\includegraphics[width=8.8cm]{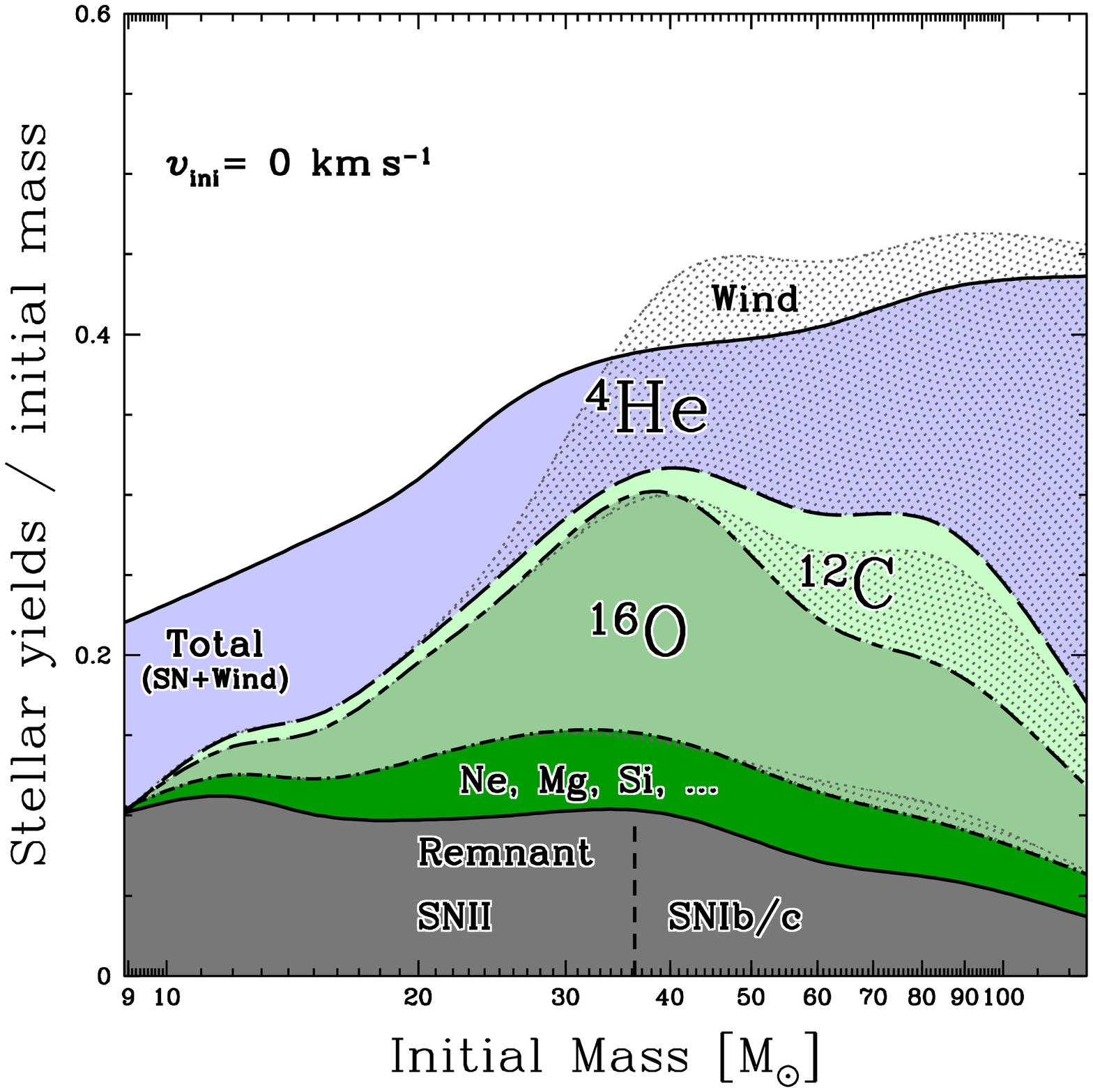}\includegraphics[width=8.8cm]{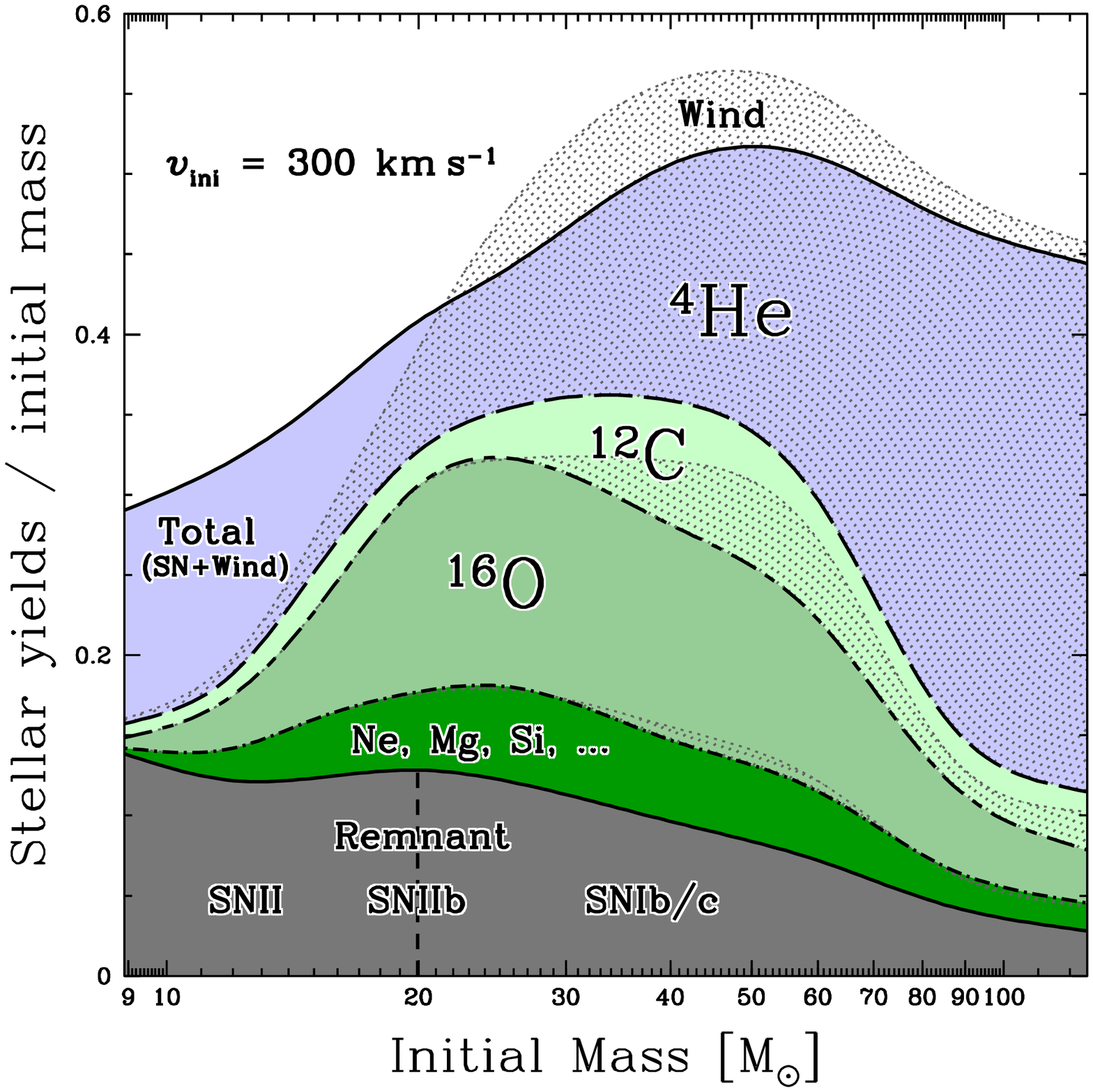}
\caption{Stellar yields divided by the initial mass, 
$p^{\rm{tot}}_{im}$, as a function of the initial mass
for the non--rotating 
 (left) and rotating  (right) 
models at solar metallicity. 
The different total yields (divided by $m$) are shown as 
piled up on top of 
each other and are not overlapping. $^4$He
yields are delimited by the top solid and long dashed lines (top
shaded area),
$^{12}$C yields by the long dashed and short--long dashed lines, 
$^{16}$O  yields by the short--long dashed and dotted--dashed lines and 
the rest of metals by the dotted--dashed and bottom solid lines. The
bottom solid line also represents the mass of the remnant 
($M^{\rm{int}}_{\rm{rem}}/m$). 
The corresponding SN explosion type is also given.
The wind contributions  are superimposed on these total yields for the
same elements between their bottom limit and the dotted line above
it. Dotted areas help quantify the fraction of the total yields
due to winds. Note that for
$^{4}$He, the total yields is smaller than the wind yields due to
negative SN yields (see text). Preliminary results for masses equal to 9, 85 and
120 $M_\odot$ were used in this diagram \citep[see][]{thesis}.}
\label{yres}
\end{figure*}
\begin{figure*}[!tbp]
\centering
\includegraphics[width=8.8cm]{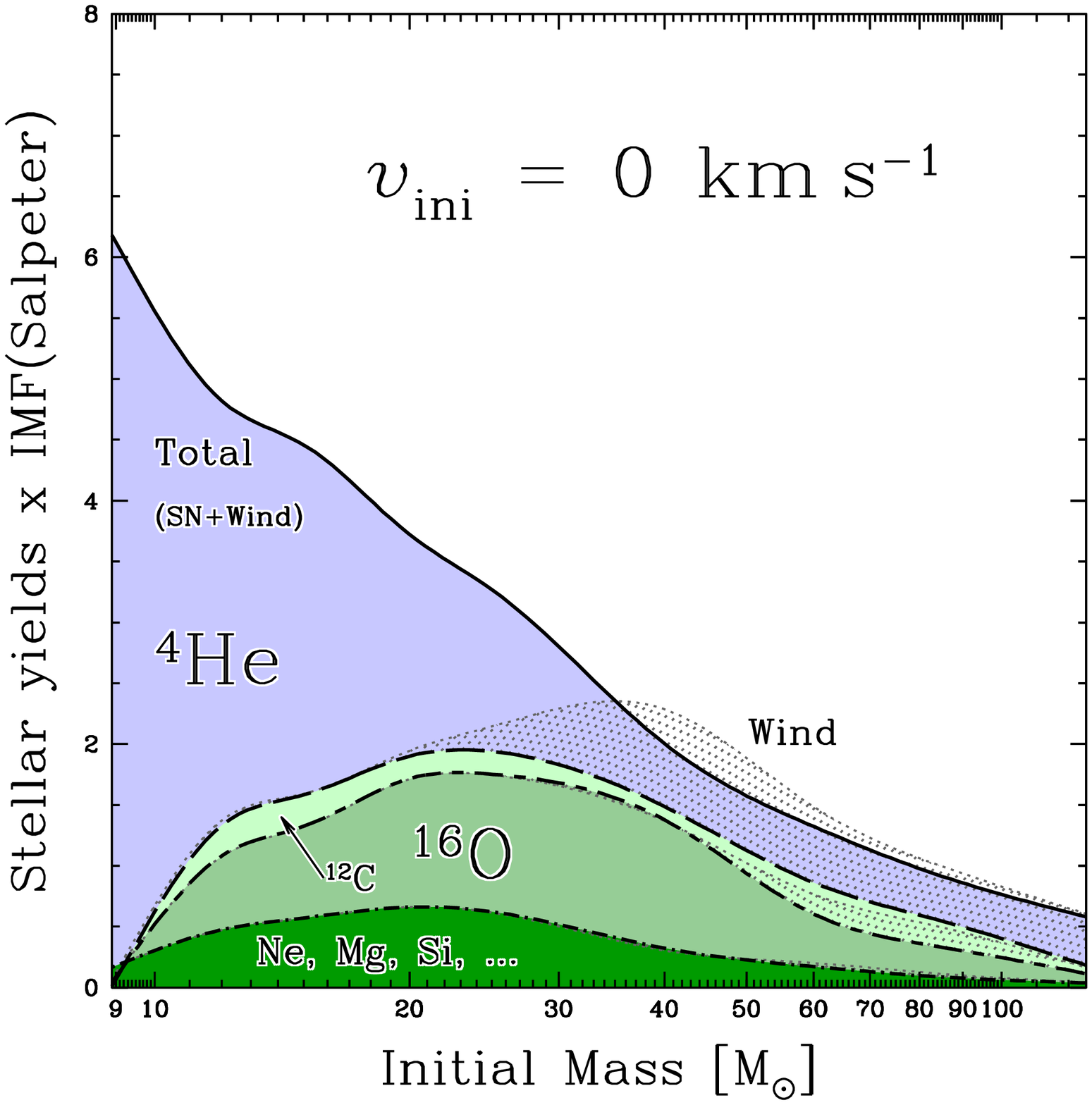}\includegraphics[width=8.8cm]{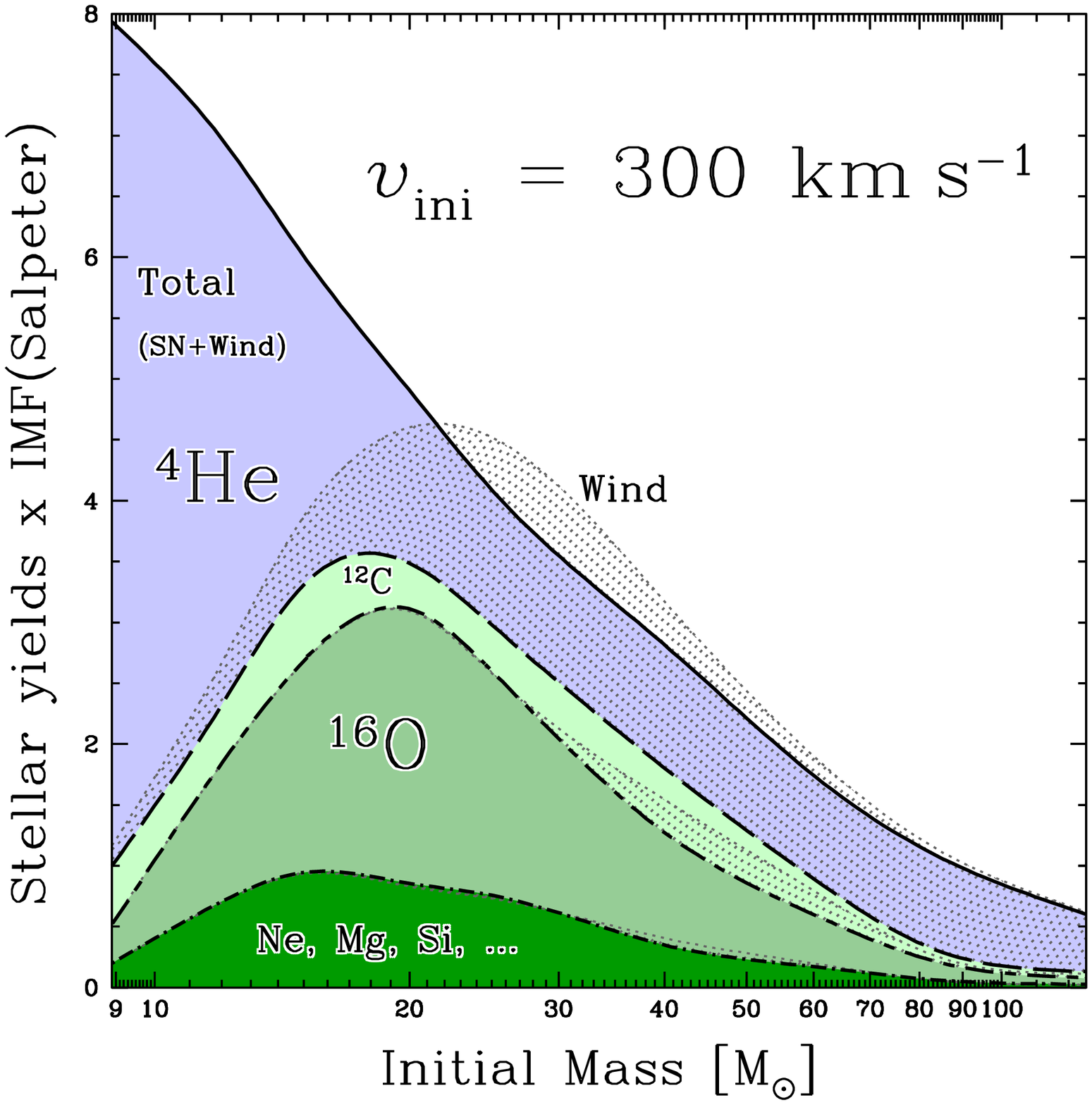}
\caption{Product of the stellar yields, $mp^{\rm{tot}}_{im}$ by 
Salpeter's
IMF (multiplied by a arbitrary constant: 1000 x $M^{-2.35}$), 
as a function of the initial mass
for the non--rotating (left) and rotating 
 (right) 
 models at solar metallicity. The different shaded areas correspond 
 from top
to bottom to $mp^{\rm{tot}}_{im}$ x 1000 x $M^{-2.35}$ for $^{4}$He, $^{12}$C, $^{16}$O and
the rest of the heavy elements. The dotted areas show for 
$^{4}$He, $^{12}$C and  $^{16}$O the wind contribution. Preliminary results for masses equal to 9, 85 and
120 $M_\odot$ were used in this diagram \citep[see][]{thesis}.}
\label{yimf}
\end{figure*}

\subsection{Comparison between the wind and pre--SN contributions}
The total stellar yields, 
$mp^{\rm{tot}}_{im}=mp^{\rm{pre-SN}}_{im} + mp^{\rm{wind}}_{im}$  
\citep[to be used for chemical evolution models using
Eq. 2 from][]{AM92}, are presented in Tables \ref{ytot} and \ref{ytotb}.
What is the relative importance of the wind and pre--SN contributions? 
Figure \ref{yres} displays the total stellar yields 
divided by the initial mass of the star, $p^{\rm{tot}}_{im}$, 
as a function of its initial mass, $m$, for the non--rotating (left) 
and rotating (right) models. 
The different shaded areas correspond from top
to bottom to $p^{\rm{tot}}_{im}$ for $^{4}$He, $^{12}$C, $^{16}$O and
the rest of the heavy elements. The fraction of the star locked in the
remnant as well as the expected explosion type are shown at the bottom. 
The dotted areas show the wind contribution for 
$^{4}$He, $^{12}$C and  $^{16}$O. 

For $^4$He (and other H--burning products like $^{14}$N), the wind
contribution increases with mass and dominates for $M \gtrsim 22
M_{\sun}$  for rotating stars and $M \gtrsim 35
M_{\sun}$  for non--rotating stars, i. e. for the stars which enter
the WR stage. As said earlier, for very massive
stars, the SN contribution is negative and this is why
$p^{\rm{tot}}_{^4\rm{He} m}$ is smaller than 
$p^{\rm{wind}}_{^4\rm{He} m}$.
In order to eject He--burning products, a star must not only become a WR 
star but must also become a WC star. 
Therefore for $^{12}$C, the wind contributions only start to be 
significant above
the following approximative mass limits: 
30 and 45 $M_{\sun}$ for rotating and
non--rotating models respectively. 
Above these mass limits, the
contribution from the wind and the pre--SN are of similar importance.
Since at solar metallicity, no WO star is produced \citep{ROTXI}, for 
$^{16}$O, as for heavier elements, the wind contribution remains very small.

\subsection{Comparison between rotating and non--rotating models}
For H--burning products, the yields of the rotating models are usually 
higher than those of
non--rotating models. This is due to larger cores and larger mass loss.
Nevertheless, between about 15 and 25 $M_{\sun}$, the rotating yields
are smaller. This is due to the fact that the winds do not expel many
H--burning products yet and more of these products are burnt
later in the
pre--supernova evolution (giving negative SN yields). 
Above 40 $M_{\sun}$,
rotation clearly increases the yields of $^4$He.

Concerning He--burning products, below 30 $M_{\sun}$, most of the 
$^{12}$C comes for the pre--SN contribution.
In this mass range, rotating models having larger cores also have larger
yields (factor 1.5--2.5). 
We notice a similar dependence on the initial mass for the yields of non--rotating
models as for the yields of rotating models, but shifted to
higher masses.
Above 30 $M_{\sun}$,
where mass loss dominates, the yields from the rotating models are
closer to those of the non--rotating models. 
The situation for $^{16}$O and metallic yields is similar to carbon. Therefore 
{\bf$^{12}$C, $^{16}$O and the total metallic
yields are larger for our rotating models compared to our non--rotating
ones by a factor 1.5--2.5 below 30 ${\mathbf M_\odot}$}.

Figure \ref{yimf} presents the stellar yields convolved with the Salpeter 
initial mass
function (IMF) ($dN/dM\propto M^{-2.35}$). This reduces the importance
of the very massive stars. Nevertheless, the differences between rotating and
non--rotating models remain significant, especially around $20\,M_\odot$.

\begin{table}
\caption{{\bf Total stellar yields} ($mp^{\rm{pre-SN}}_{im} + mp^{\rm{wind}}_{im}$) 
 of solar metallicity models. Continuation of Table \ref{ytot}. 
 Note that $^{20}$Ne yields are an upper limit 
and may be reduced by Ne--explosive burning and that $^{24}$Mg yields can
also be modified by neon and oxygen explosive burnings. 
See discussion in Sect. \ref{yao}.}
\begin{tabular}{l l r r r}
\hline
\hline \\
$M_{\rm{ini}}$ & $\upsilon_{\rm{ini}}$ & ($^{20}$Ne) & $^{22}$Ne & ($^{24}$Mg) \\
\hline
 12 &    0   &  1.05E-1 &  6.80E-3 &  6.75E-3       \\
 12 &  300   &  1.58E-1 &  1.55E-2 &  1.36E-2       \\
 15 &    0   &  1.10E-1 &  1.60E-2 &  6.06E-2       \\
 15 &  300   &  2.26E-1 &  3.33E-2 &  4.27E-2       \\
 20 &    0   &  4.83E-1 &  3.64E-2 &  1.28E-1       \\
 20 &  300   &  6.80E-1 &  4.26E-2 &  1.19E-1       \\
 25 &    0   &  8.41E-1 &  5.22E-2 &  1.38E-1       \\
 25 &  300   &  1.08E+0 &  2.26E-2 &  1.48E-1       \\
 40 &    0   &  1.36E+0 &  5.58E-3 &  1.52E-1       \\
 40 &  300A  &  1.42E+0 &  8.05E-2 &  1.20E-1       \\
 60 &    0   &  1.81E+0 &  1.35E-1 &  1.73E-1       \\
 60 &  300A  &  1.75E+0 &  1.74E-1 &  1.44E-1       \\
\hline
\end{tabular}
\label{ytotb}
\end{table}
\subsection{Comparison with the literature}\label{yao}
\begin{figure*}[!tbp]
\centering
\includegraphics[width=8.8cm]{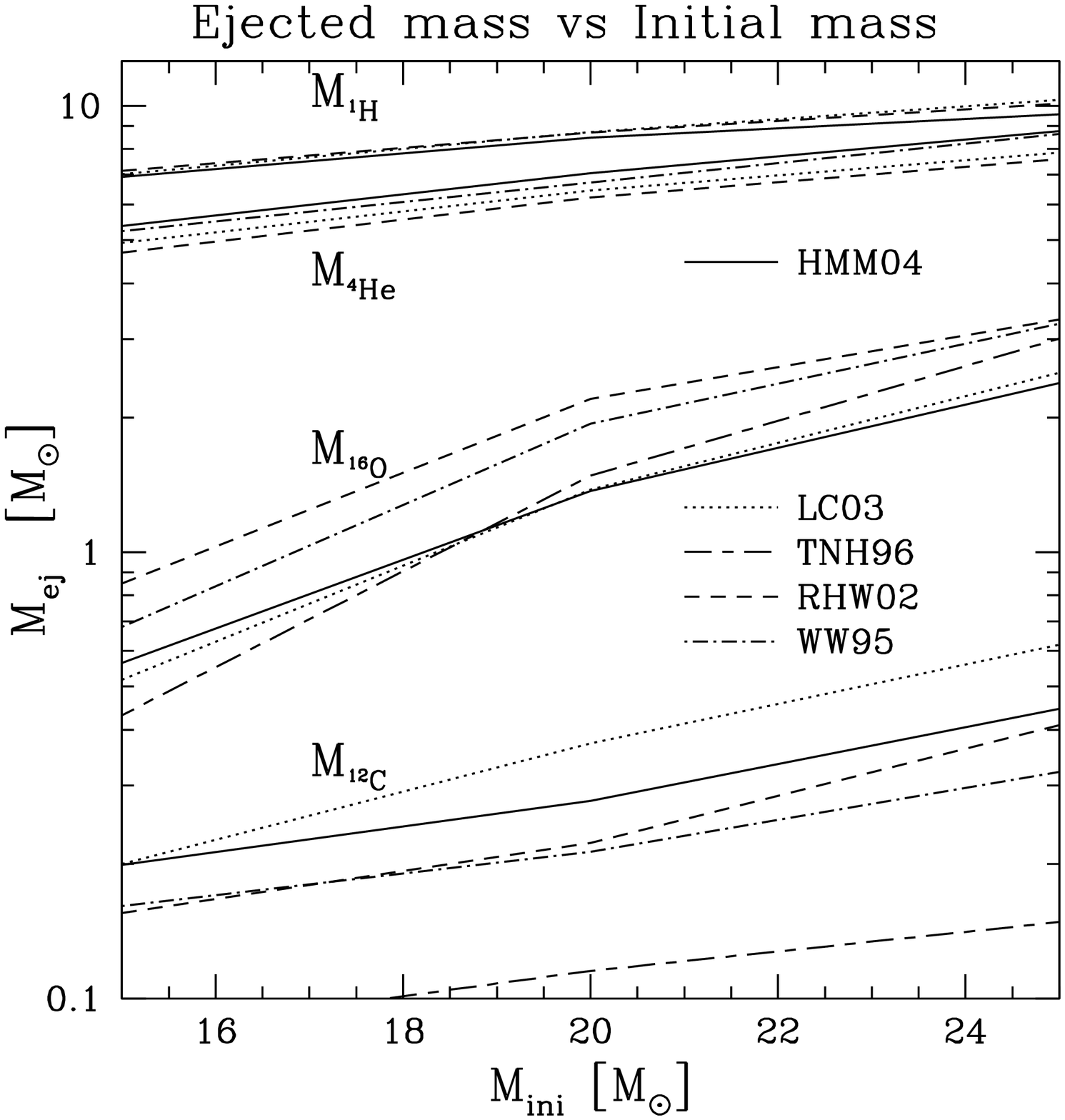}\includegraphics[width=8.8cm]{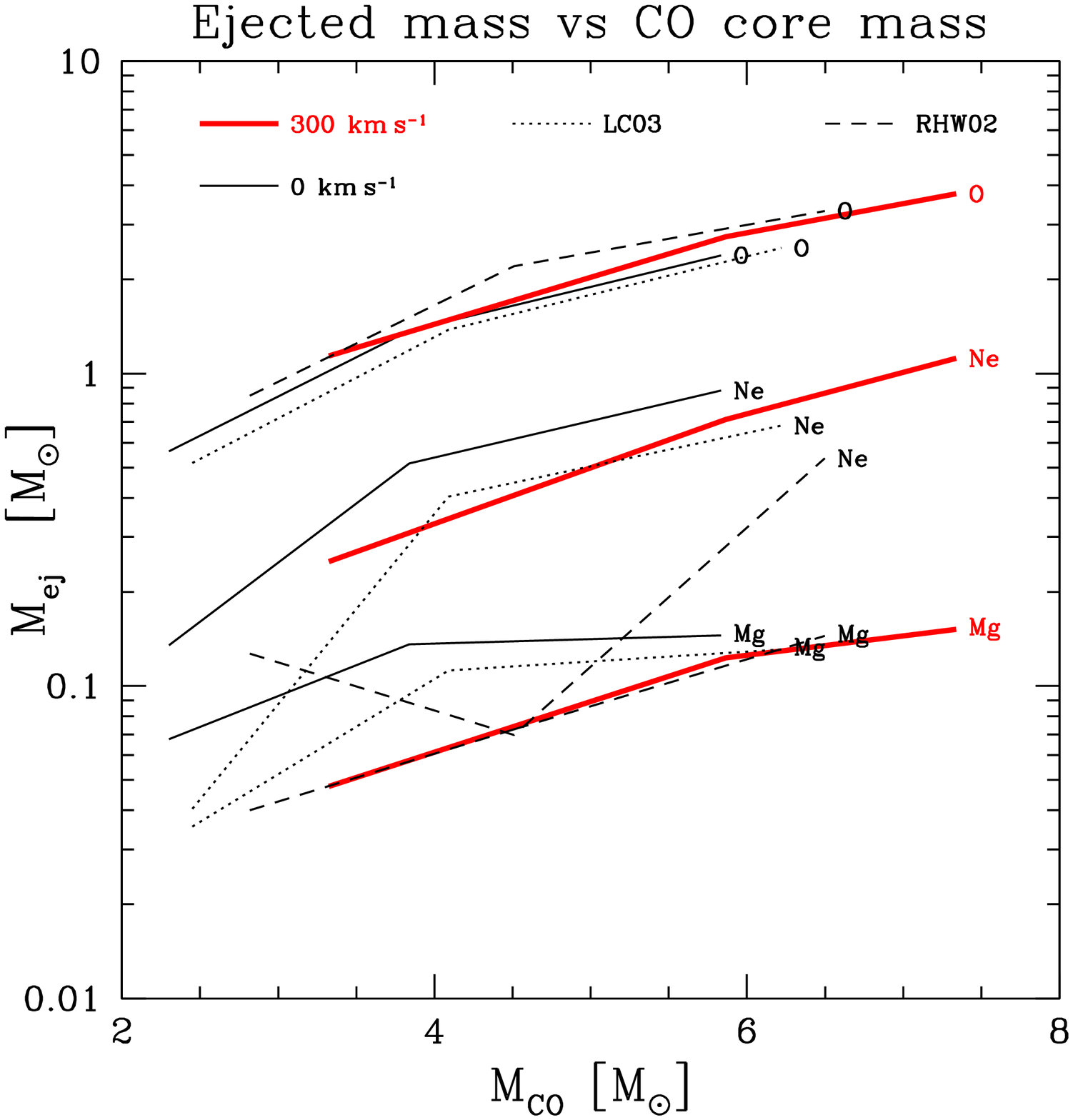}
\caption{{\it Left}: total ejected masses (EM) of 
$^1$H, $^4$He, $^{12}$C and $^{16}$O
as a function of the initial mass
for different non--rotating 
models at solar metallicity.
{\it Right}: total ejected masses (EM) of 
$^{16}$O, $^{20}$Ne and $^{24}$Mg
as a function of the CO core mass
for different 
models at solar metallicity. 
Solid lines (HMM04) represent our results,
dotted lines (LC03) show the results from \citet{LC03},
long--short dashed lines (TNH96) show the results from \citet{TNH96},
dashed lines (RHW02) represent the results from \citet{RHHW02} and
dotted--dashed (WW95) lines show the results from \citet{WW95}.
}
\label{abc}
\end{figure*}
\begin{table*}
\caption{
Initial mass and velocity, remnant mass and 
{\bf total ejected masses (EM)}
 of solar metallicity models. All masses are in
solar mass units and velocities are in km\,s$^{-1}$.
"A" in column 2 means wind anisotropy has been included in the model. Note that this
table is given for comparison with other recent publications and does not
correspond to our definition of yields (see Sect. \ref{def}).}
\begin{tabular}{l l r r r r r r r r r r r}
\hline
\hline \\
$M_{\rm{ini}}$ & $\upsilon_{\rm{ini}}$ & $M_{\rm{rem}}$ & $^1$H & $^{3}$He & $^4$He    & $^{12}$C & $^{13}$C & $^{14}$N & $^{16}$O & $^{17}$O & $^{18}$O & Z \\
\hline
 12 &    0   &  1.34 &  5.86E+0 &   1.87E-4 &  4.13E+0 &  1.23E-1 &  1.05E-3 &  4.81E-2 &  3.09E-1 &  7.75E-5 &  3.95E-4 &  6.70E-1  \\   
 12 &  300   &  1.46 &  5.03E+0 &   1.46E-4 &  4.50E+0 &  1.99E-1 &  1.51E-3 &  5.16E-2 &  4.93E-1 &  7.58E-5 &  1.91E-3 &  1.01E+0  \\   
 15 &    0   &  1.51 &  6.92E+0 &   2.02E-4 &  5.38E+0 &  1.99E-1 &  1.26E-3 &  5.76E-2 &  5.64E-1 &  6.86E-5 &  3.06E-3 &  1.19E+0  \\   
 15 &  300   &  1.85 &  5.78E+0 &   1.56E-4 &  5.19E+0 &  4.06E-1 &  1.61E-3 &  4.91E-2 &  1.14E+0 &  7.14E-5 &  4.01E-3 &  2.18E+0  \\   
 20 &    0   &  1.95 &  8.48E+0 &   2.28E-4 &  7.05E+0 &  2.77E-1 &  1.38E-3 &  6.88E-2 &  1.37E+0 &  7.94E-5 &  4.31E-3 &  2.53E+0  \\   
 20 &  300   &  2.57 &  6.69E+0 &   1.72E-4 &  6.41E+0 &  4.93E-1 &  1.73E-3 &  6.17E-2 &  2.74E+0 &  6.52E-5 &  1.89E-4 &  4.33E+0  \\   
 25 &    0   &  2.49 &  9.57E+0 &   2.22E-4 &  8.76E+0 &  4.45E-1 &  1.34E-3 &  9.79E-2 &  2.39E+0 &  8.45E-5 &  2.85E-4 &  4.19E+0  \\   
 25 &  300   &  3.06 &  7.57E+0 &   1.89E-4 &  8.18E+0 &  8.12E-1 &  1.86E-3 &  9.70E-2 &  3.76E+0 &  6.87E-5 &  7.51E-4 &  6.19E+0  \\   
 40 &    0   &  4.02 &  1.36E+1 &   3.40E-4 &  1.29E+1 &  7.72E-1 &  1.78E-3 &  1.73E-1 &  6.47E+0 &  9.01E-5 &  2.69E-4 &  9.37E+0  \\   
 40 &  300A  &  3.85 &  9.11E+0 &   2.14E-4 &  1.58E+1 &  3.24E+0 &  1.67E-3 &  2.02E-1 &  5.71E+0 &  6.73E-5 &  1.93E-4 &  1.12E+1  \\   
 60 &    0   &  4.30 &  1.91E+1 &   4.85E-4 &  2.22E+1 &  4.13E+0 &  2.28E-3 &  2.84E-1 &  7.01E+0 &  1.14E-4 &  3.68E-4 &  1.41E+1  \\   
 60 &  300A  &  4.32 &  1.29E+1 &   2.56E-4 &  2.81E+1 &  4.70E+0 &  2.18E-3 &  3.57E-1 &  6.94E+0 &  7.86E-5 &  1.94E-4 &  1.46E+1  \\   
\hline
\end{tabular}
\label{etot}
\end{table*}
We compare here the yields of the non--rotating models 
with other authors. For this purpose, 
the ejected masses, $EM$, defined by Eq. \ref{emdef} 
in Sect. \ref{def}, are presented in
Tables \ref{etot} and \ref{etotb}. Figure \ref{abc} shows the comparison
with four other calculations: \citet{LC03} (LC03), 
\citet{TNH96} (TNH96), \citet{RHHW02} (RHW02) and \citet{WW95} (WW95).
For LC03, we chose the remnant masses that are closest to
ours (models 15D, 20B, 25A). 
The uncertainties related to convection and the
$^{12}$C$(\alpha,\gamma)^{16}$O reaction are dominant. 
Therefore, before we compare our results with other models, 
we briefly mention here which treatment of
convection and $^{12}$C$(\alpha,\gamma)^{16}$O rate other authors use:
\begin{itemize}
\item \citet{LC03} use Schwarzschild criterion (except for the H convective shell
that forms at the end of core H--burning, where Ledoux criterion is used) 
for convection without 
overshooting. For $^{12}$C$(\alpha,\gamma)^{16}$O, they use the rate
of \citet{K02} (K02).
\item \citet{TNH96} use Schwarzschild criterion for convection without 
overshooting. For $^{12}$C$(\alpha,\gamma)^{16}$O, they use the rate
of \citet{CF85} (CF85).
\item \citet{WW95} use Ledoux criterion for convection with 
semiconvection. They use a relatively large diffusion coefficient
to model semiconvection. Moreover non--convective zones
immediately adjacent to convective regions are slowly mixed
over the order of a radiation diffusion time scale to 
approximately allow for the effects of convective overshoot.
For $^{12}$C$(\alpha,\gamma)^{16}$O, they use the rate
of \citet{CF88} (CF88) multiplied by 1.7.
\item \citet{RHHW02} use Ledoux criterion for convection with 
semiconvection. They use the same 
method as WW95 for semiconvection. 
For $^{12}$C$(\alpha,\gamma)^{16}$O, they use the rate
of \citet{BU96} (BU96) multiplied by 1.2.
\item In this paper (HMM04), we use Schwarzschild criterion 
for convection with overshooting. For 
$^{12}$C$(\alpha,\gamma)^{16}$O, we use the rate
of \citet{NACRE} (NACRE).
\end{itemize}

A comparison of the different reaction rates and treatment of convection
 is presented in \citet{psn04a}. 
The comparison of the ejected masses is shown in Fig. \ref{abc} for 
masses between 15 and 25 $M_\odot$.
The $^{4}$He and $^{16}$O yields are larger when respectively 
the helium and carbon--oxygen cores are larger. 
This can be seen by comparing our models with those of RHW02 and LC03 
(Fig. \ref{abc} and respective tables of core masses). 

For $^{12}$C
yields, the situation is more complex because the larger the cores, the
larger the central temperature and the more efficient the 
$^{12}$C$(\alpha,\gamma)^{16}$O reaction.
If we only consider the effect of this reaction we have that the larger 
the rate, the smaller the $^{12}$C
abundance at the end of He--burning and the smaller the corresponding
yield (and the larger the $^{16}$O yield). This can be seen in Fig.
\ref{abc} by comparing
our $^{12}$C and $^{16}$O yields with those of LC03 
(we both use Schwarzschild criterion). Indeed the NACRE
rate is larger than the K02 one so our $^{12}$C yield is smaller. 
THN96 (who also use Schwarzschild criterion) 
using the rate of \citet{CF85} 
which is even larger, obtain an even smaller $^{12}$C yield. 
When both the convection treatment and the
$^{12}$C$(\alpha,\gamma)^{16}$O rate are different, the comparison 
becomes 
more complicated. Nevertheless, within the model uncertainties, the
yields of various models agree. In fact, the uncertainties are reduced 
when we use the CO core mass instead of the initial mass in order to 
compare the 
results of different groups. Fig. \ref{abc} (right) shows the small
uncertainty for $^{16}$O in relation to the CO core mass. This confirms
the relation $M_{\rm{CO}}$--yields($^{16}$O) and shows that this
relation holds for models of different groups and for models of
non--rotating and rotating stars.

We calculated the pre--SN yields at the end of Si--burning. Therefore, 
the yields of $^{20}$Ne and $^{24}$Mg may still be affected by 
explosive neon and oxygen burnings.
$^{20}$Ne yields are upper
limits due to the possible destruction of this element by
explosive Ne--burning. Figure \ref{abc} (right) shows that our results lies above
the results of other groups but that the difference is as small as differences between
the results of the other groups. $^{24}$Mg yields are also close to the results 
of other groups who included explosive burnings in their calculations. 
By comparing our results for $^{20}$Ne and $^{24}$Mg with the other groups 
mentioned above, we see that the difference between our results and the results
of other groups is as small as the differences between the 19, 20 and 21 $M_\odot$
models of \citet{RHHW02} and differences between for example \citet{RHHW02} and 
\citet{LC03}. This means that our yields for these two elements are 
good approximations even though explosive burning was 
not followed in this calculation. For $^{24}$Mg, it is interesting to
note that rotation increases significantly the yields only for the 12 
$M_\odot$ models and that, in general, rotation slightly decreases the 
$^{24}$Mg yields in the massive star range
(see Table \ref{ytotb} and Fig. \ref{abc} right).
This point is interesting for chemical evolution of galaxies since
it goes in the same direction as observational constraints \citep{CC04}.

For $^{17}$O
yields, all recent calculations agree rather well and differ from the
WW95 results because of the change in the reaction rates 
\citep[especially $^{17}$O$(p,\alpha)^{14}$N, see][]{APB96}.
$^{18}$O and $^{22}$Ne are produced by $\alpha$--captures on $^{14}$N. 
As said in Sect. \ref{def}, $^{22}$Ne is not
followed during the advanced stages and we had to use a special
calculation for its yield. Our $^{22}$Ne values 
are nevertheless very close to other calculations \citep[see][]{thesis}.
\begin{table}
\caption{{\bf Total ejected masses (EM)}
 of solar metallicity models. Continuation of Table \ref{etot}.}
\begin{tabular}{l l r r r}
\hline
\hline \\
$M_{\rm{ini}}$ & $\upsilon_{\rm{ini}}$ & ($^{20}$Ne) & $^{22}$Ne & ($^{24}$Mg)  \\
\hline
 12 &    0   &  1.24E-1 &  8.37E-3 &  1.27E-2   \\   
 12 &  300   &  1.77E-1 &  1.71E-2 &  1.87E-2   \\   
 15 &    0   &  1.35E-1 &  1.80E-2 &  6.75E-2   \\   
 15 &  300   &  2.50E-1 &  3.53E-2 &  4.77E-2   \\   
 20 &    0   &  5.16E-1 &  3.90E-2 &  1.36E-1   \\   
 20 &  300   &  7.12E-1 &  4.52E-2 &  1.23E-1   \\   
 25 &    0   &  8.82E-1 &  5.55E-2 &  1.45E-1   \\   
 25 &  300   &  1.12E+0 &  2.59E-2 &  1.52E-1   \\   
 40 &    0   &  1.42E+0 &  1.08E-2 &  1.58E-1   \\   
 40 &  300A  &  1.48E+0 &  8.58E-2 &  1.25E-1   \\   
 60 &    0   &  1.91E+0 &  1.43E-1 &  1.79E-1   \\   
 60 &  300A  &  1.86E+0 &  1.82E-1 &  1.50E-1   \\   
\hline
\end{tabular}
\label{etotb}
\end{table}
\section{Conclusion}
We calculated a new set of stellar yields of rotating
stars at solar metallicity covering the  massive star 
range (12--60 $M_{\sun}$). We used
for this purpose the latest version of the Geneva stellar evolution code
described in \citet{psn04a}. 
We present the separate contribution to stellar yields by winds and
supernova explosion. 
For the wind contribution, our rotating models have larger yields than
the non--rotating ones because of the extra mass loss and mixing due to
rotation. 
For the SN yields, we followed the evolution and nucleosynthesis until 
core silicon burning.
Since we did not model the SN explosion and
explosive nucleosynthesis, we present pre--SN yields for elements 
lighter than 
$^{28}$Si,
which depend mostly on the evolution prior to Si--burning.
Our results for the non--rotating models correspond very well
to other calculations and differences can be understood in the light
of the treatment of convection and the rate used for
$^{12}$C$(\alpha,\gamma)^{16}$O. This assesses the accuracy of our
calculations and assures a safe basis for the yields of our rotating models.

For the pre--SN yields and for masses below $\sim 30\,M_{\sun}$, 
rotating models have larger yields. 
The $^{12}$C and $^{16}$O yields are increased by a factor of 1.5--2.5 by
rotation in the present calculation. 
When we add the two contributions, the yields of most
heavy elements are larger for rotating models below $\sim 30\,M_{\sun}$. 
Rotation increases
the total metallic yields by a factor of 1.5--2.5. 
As a rule of thumb, the yields of a rotating 20 $M_\odot$ star are similar
to the yields of a non--rotating 30 $M_\odot$ star, at least for the light
elements considered in this work.
When
mass loss is dominant (above $\sim 30\,M_{\sun}$) our rotating and
non--rotating models give similar yields for heavy elements.
Only the yields of H--burning products are increased by rotation in the 
very massive star range. 

%


\bibliographystyle{aa}

\end{document}